\documentclass[final,3p,times,number,compress]{elsarticle}
\usepackage{amssymb}
\usepackage{lipsum}
\usepackage{bm}
\usepackage{verbatim}
\usepackage{empheq}

\usepackage{amsmath,amsfonts,amsthm,amssymb}
\usepackage{graphics}
\usepackage{cancel}
\usepackage{multirow}
\usepackage{longtable}
\usepackage{color}
\usepackage{xcolor}
\usepackage[normalem]{ulem}
\usepackage{dsfont}

\journal{Physics Letters B}

\newcommand{\beq}{\begin{equation}}
\newcommand{\eeq}{\end{equation}}

\newcommand{\bea}{\begin{eqnarray}}
\newcommand{\eea}{\end{eqnarray}}

\newcommand{\bel}[1]{\begin{eqnarray}\label{#1}}
\newcommand{\eel}{\end{eqnarray}}

\newcommand{\EQ}[1]{Eq.~(\ref{#1})}
\newcommand{\EQn}[1]{(\ref{#1})}

\newcommand{\EQSTWOn}[2]{(\ref{#1})~and~(\ref{#2})}

\newcommand{\EQSM}[2]{Eqs.~(\ref{#1})--(\ref{#2})}
\newcommand{\EQSMn}[2]{(\ref{#1})--(\ref{#2})}

\newcommand{\CITn}[1]{\citep{#1}} 

\newcommand{\dd}{\mathrm{d}}

\newcommand{\Bv}{{\boldsymbol B}} 
 
\newcommand{\Ev}{{\boldsymbol E}} 
\newcommand{\Pv}{{\boldsymbol P}}

\newcommand{\av}{{\boldsymbol a}} 
\newcommand{\bv}{{\boldsymbol b}} 
 
\newcommand{\ev}{{\boldsymbol e}}

\newcommand{\nv}{{\boldsymbol n}} 

\newcommand{\vv}{{\boldsymbol v}}
\newcommand{\pv}{{\boldsymbol p}}

\newcommand\sigv{{\boldsymbol \sigma}}

\newcommand{\trt}{{\rm tr_2}}
\newcommand{\trf}{{\rm tr_4}}

\usepackage{amssymb}

\usepackage{graphicx}
\usepackage{mathtools}
\usepackage{subfigure}
\usepackage{bbold}
\usepackage{yfonts}
\usepackage[colorlinks=true,linktocpage=true,linkcolor=blue,citecolor=blue]{hyperref}
\usepackage{placeins}
\usepackage{bm} 
\usepackage{nicefrac}
\usepackage{slashed}
\usepackage{marginnote}

\newcommand{\f}[2]{\frac{#1}{#2}}
\newcommand{\onehalf}{{\nicefrac{1}{2}}}

\def\ubarrp{{\bar u}_r(p)}
\def\ubarsp{{\bar u}_s(p)}
\def\usp{u_s(p)}
\def\urp{u_r(p)}

\def\vbarrp{{\bar v}_r(p)}
\def\vbarsp{{\bar v}_s(p)}
\def\vsp{v_s(p)}
\def\vrp{v_r(p)}

\def\fplusrsxp{f^+_{rs}(x,p)}

\def\fminusrsxp{f^-_{rs}(x,p)}

\begin{document}

\begin{frontmatter}



\title{Fermi--Dirac Wigner function for massive spin-$\nicefrac{1}{2}$ particles in local equilibrium}


\author[first]{Sudip Kumar Kar}
\affiliation[first]{organization={Institute of Theoretical Physics, Jagiellonian University},
            addressline={ul. St. \L ojasiewicza 11}, 
            city={Krak\'ow},
            postcode={30-348}, 
            %state={},
            country={Poland}
            }
\ead{sudip.kar@doctoral.uj.edu.pl}
\author[first]{Valeriya Mykhaylova}
\ead{valeriya.mykhaylova@uj.edu.pl}
\begin{abstract}
A recently proposed Boltzmann local equilibrium Wigner function for massive spin-$\nicefrac{1}{2}$ particles is generalized to the case of Fermi--Dirac  statistics. The resulting formula ensures the correct normalization of the mean polarization vector and reproduces the generalized thermodynamic relations with spin that were obtained in earlier studies. Moreover, we show that the macroscopic currents constructed from the Fermi--Dirac Wigner function can be obtained as derivatives of a suitably defined generating function with respect to the Lagrange multipliers (temperature, hydrodynamic flow, and chemical potentials). The identified generating function also indicates that the underlying framework can be classified as a divergence-type theory.
\end{abstract}



\begin{keyword}
relativistic kinetic theory \sep relativistic hydrodynamics \sep spin polarization \sep Fermi--Dirac statistics



\end{keyword}

\end{frontmatter}

%\tableofcontents


\section{Introduction}
\label{introduction}

Nonzero spin polarization of particles produced in heavy-ion collisions has been a subject of great interest ever since its predictions~\CITn{Liang:2004ph, Liang:2004xn,PhysRevC.76.044901,Voloshin:2004ha,Becattini:2007nd,Becattini:2007sr,Becattini:2013fla,Becattini:2013vja,Becattini:2016gvu} and subsequent successful \mbox{measurements~\CITn{STAR:2017ckg, STAR:2018gyt, STAR:2019erd, ALICE:2019aid,STAR:2025jhp,STAR:2022fan,Li:2025awj,STAR:2025jwc,Shen:2024rdq}.} The observed global polarization can be interpreted as evidence for the thermalization of spin degrees of freedom,
suggesting the incorporation of spin into existing theoretical frameworks. In particular, it should be included in relativistic hydrodynamics~\cite{Florkowski:2017ruc}, which has established itself as a standard tool for modeling heavy-ion collisions~\cite{Romatschke:2017ejr,Heinz:2013th}. 

In the context of relativistic hydrodynamics and the underlying relativistic kinetic theory, the form of the local equilibrium distribution function for spin-$\onehalf$ particles is significantly relevant. Its first version was introduced in~\CITn{Becattini:2013fla}; however, it was later found that this form violates the normalization of the mean spin polarization vector~\CITn{Florkowski:2017dyn}, which should be bounded from above. One way to address this problem was through an approach based on the classical description of spin~\CITn{Florkowski:2018fap}. Only very recently has an alternative formula for the local equilibrium Wigner function with a quantum description of spin been proposed~\CITn{Bhadury:2025boe}. This formulation solves the issue with the mean spin polarization and provides a suitable framework of perfect spin hydrodynamics, including consistent thermodynamic relations and a divergence-type structure of the equations of motion that ensures nonlinear causality and stability.

In this work, we relax the final key assumption used \mbox{in~\CITn{Bhadury:2025boe}} by replacing the Boltzmann statistics, which is valid for dilute systems, with Fermi--Dirac statistics. Phenomenologically, is it necessary to develop a Fermi--Dirac form of the spin distribution function, as such an object enables analysis of systems at relatively lower temperatures and high baryon densities, where the spin polarization effects are found to be the most pronounced~\cite{PhysRevC.100.014908, PhysRevC.101.064908,Deng:2021miw,PhysRevC.104.L041902}. 

We show that the Fermi--Dirac local equilibrium Wigner function defined below can be used to construct microscopic currents that satisfy generalized thermodynamic relations for particles with spin, in agreement with earlier \mbox{findings~\CITn{Florkowski:2024bfw,Drogosz:2024gzv}}. We additionally demonstrate that one can define a generating function for the Fermi--Dirac case, whose gradients with respect to the Lagrange multipliers yield conserved currents. This indicates that our framework can be cast into the scheme of a divergence-type theory.

{\it Notation and conventions} --- Throughout the paper, we adopt the natural units, \mbox{$\hbar = c = k_{\rm B} = 1$}. The metric tensor is used in the ``mostly minuses'' convention, \mbox{$g_{\mu\nu} = \textrm{diag}(+1,-1,-1,-1)$}, thus the scalar product of two four-vectors $a$ and $b$ is defined as \mbox{$a \cdot b = a^0 b^0 - \av \cdot \bv$}, with the three-vectors indicated in bold. We also define \mbox{${\tilde a}^{\alpha\beta} \equiv (\onehalf) \,\epsilon^{\alpha\beta\gamma\delta} a_{\gamma \delta}$} to be the dual tensor to $a^{\alpha\beta}$, with the sign of the Levi-Civita symbol $\epsilon^{\alpha\beta\gamma\delta}$ fixed by the condition \mbox{$\epsilon^{0123} =-\epsilon_{0123} = +1$}. The traces over the spin and spinor indices are denoted by $\trt$ and $\trf$, respectively. Finally, we use the Feynman slash notation, $\slashed{a} = a_\mu \gamma^\mu$, involving the Dirac matrices $\gamma^\mu$.

The paper is organized as follows: In Sec.~\ref{sec:spin_FD}, we examine the main components of the recently proposed local equilibrium Wigner function for spin-$\nicefrac{1}{2}$ particles obeying Boltzmann distribution and extend it to the case of Fermi--Dirac statistics. Section~\ref{sec:2x2} describes the transition from the four-by-four spinor density matrices to the two-by-two spin density matrices, which determine the magnitude of the mean spin polarization vector. We show that the direction of the latter coincides, as expected, with that of the orbital angular momentum of the rigidly rotating system in global equilibrium. In Sec.~\ref{sec:currents}, we introduce the macroscopic currents and demonstrate that they satisfy the generalized first law of spin thermodynamics. In Sec.~\ref{sec:GenFun} we describe the divergence type framework of perfect spin hydrodynamics and give the corresponding generating function for all the conserved currents in our approach.
We summarize and conclude in Sec.~\ref{sec:summary}. The two Appendices provide the additional details of our computations.
\medskip
%
%%%%%%%%%%%%%%%%%%%%%%%%%%%%%
\section{Transition from the Boltzmann to the Fermi--Dirac case}
\label{sec:spin_FD}

Recently, the local equilibrium Wigner function for spin-$\nicefrac{1}{2}$ particles was introduced in~\cite{Bhadury:2025boe} in the form
\begin{align}
    W^\pm(x,k)\!=\!\frac{1}{4 m}  \int \!\dd P\,
\delta^{(4)}(k\!\mp\!p) 
(\slashed{p}\!\pm\! m) X^\pm(x,p)
(\slashed{p}\!\pm\! m). 
\label{eq:Wpm}
\end{align}
Here the superscript $+ \, (-)$ refers to particles (antiparticles), $m$~denotes the particle mass, and $p = (E_p, \pv)$ is the on-shell particle four-momentum with $p^0 = E_p = \sqrt{m^2+\pv^2}$. The momentum integration measure is given by \mbox{$\dd P = \dd^3p/((2\pi)^3 E_p)$}. The quantity $X^\pm$ is the four-by-four spinor density matrix,
\begin{align}
    X^{\pm}(x,p) =  \exp\left[\pm \xi(x) - \beta (x) \cdot  p + \gamma_5 \slashed{a}(x,p) \right], 
    \label{eq:Xpm}
\end{align}
where $\xi = \mu/T$ is the ratio of the baryon chemical potential to the temperature and $\beta^\mu= u^\mu/T$ involves the fluid four-velocity normalized to unity, $u^\mu = \gamma (1, \vv )$, with $\gamma$ being the Lorentz factor. The space-like four vector $a_\mu$ is defined as~\cite{Bhadury:2025boe,Florkowski:2019gio}
\begin{align}
    a_\mu(x,p) = -\frac{1}{2 m} {\tilde \omega}_{\mu\nu}(x)p^\nu,
    \label{eq:amu}
\end{align}
where ${\tilde \omega}_{\mu\nu}$ represents the dual of the spin polarization tensor~$\omega_{\mu\nu}$. The latter is specified by the ratio of the spin chemical potential, $\Omega_{\mu\nu}$, to the temperature, $\omega_{\mu\nu} = \Omega_{\mu\nu}/T$ (note that both $\xi$ and $\omega_{\mu\nu} $ are dimensionless in natural units). Following earlier works~\cite{Florkowski:2017dyn,Bhadury:2025boe,Florkowski:2019gio}, the spin polarization tensor $\omega_{\mu\nu}$ can be represented by its electriclike and magneticlike components, $\ev$ and~$\bv$, in exactly the same way as the Faraday tensor $F_{\mu\nu}$ in classical electrodynamics is written in terms of the electric and magnetic fields, $\Ev$ and $\Bv$. More physical insight can be gained by considering the tensors in the particle rest frame (PRF). Denoting the quantities in this frame by an asterisk, we find
\begin{align}
    a^{\mu}_{*} = (0, \av_*) = - \frac{\bv_*}{2}  \, , \qquad -a^2 = \av_*^2 > 0 \, .
    \label{eq:amustar}
\end{align}
Hence, it is clear that the spin polarization effects in equilibrium are governed by the magnetic component of the spin polarization tensor in PRF, analogous to how the orientation of magnetic moments is determined by the magnetic field (we comment further on this relation below).

Since $a^2<0$, it is convenient to rewrite the exponential of \mbox{$\gamma_5 \slashed{a}$} in \EQn{eq:Xpm} as 
\begin{align}
\exp\!\left( \gamma_5 \slashed{a} \right) = \cosh\!\sqrt{-a^2}
+ \frac{\gamma_5 \slashed{a}}{\sqrt{-a^2}} \sinh\!\sqrt{-a^2},
\end{align}
and use the definitions of the hyperbolic functions to express the spinor density matrix in the form
\begin{align}
\begin{split}
    X^\pm_{\alpha \beta}=& \frac{1}{2}\Bigg[\left(e^{\pm \xi-\beta\cdot p+\sqrt{-a^2}} + e^{\pm \xi-\beta\cdot p-\sqrt{-a^2}}\right)\,\delta_{\alpha\beta}  %\\
 +   \frac{(\gamma_5 \slashed{a})_{\alpha\beta}}{\sqrt{-a^2}} \,\left(e^{\pm \xi-\beta\cdot p+\sqrt{-a^2}} - e^{\pm \xi-\beta\cdot p-\sqrt{-a^2}}\right) \Bigg]. \label{eq:XBoltzm}
 \end{split}
\end{align}
Here, the subscripts $\alpha,\beta$ denote the spinor indices, which run from 1 to 4, and each exponential function plays the role of the Boltzmann distribution in the spin-extended phase space. Therefore, we introduce a compact notation
\begin{align}
    g_{\rm{B}}^{\pm \pm}(x,p)=\exp\left[\pm \xi(x)-\beta(x)\cdot p\pm\sqrt{-a^2(x,p)}\right], \label{eq:Bolt-short}
\end{align}
where the first $\pm$ distinguishes between particles ($+$) and antiparticles ($-$), while the second refers to the spin orientation, ($+$) for the direction aligned with $\av_*$ in PRF, and ($-$) for the opposite direction. Using this definition, \EQ{eq:XBoltzm} can be rewritten as
\begin{align}
    X^\pm_{\alpha \beta}&= \frac{1}{2}\left[\left( g_{\rm{B}}^{\pm +} +  g_{\rm{B}}^{\pm -}\right)\,\delta_{\alpha\beta} + \frac{(\gamma_5 \slashed{a})_{\alpha\beta}}{\sqrt{-a^2}} \,\left( g_{\rm{B}}^{\pm +} -  g_{\rm{B}}^{\pm -}\right) \right].
\end{align}
At this point, it is natural to generalize our Boltzmann-statistics result to the Fermi--Dirac case. This is accomplished simply by replacing the Boltzmann distribution function, $g_{\rm{B}}^{\pm \pm}(x,p)$,  with the Fermi--Dirac one,
%.
\begin{align}
     g_{\rm{FD}}^{\pm \pm}(x,p)=\left[ \exp\left({\mp \xi(x)+\beta(x)\cdot p\mp\sqrt{-a^2(x,p)}}\right)+1 \right]^{-1},
     \label{eq:FD-short}
\end{align}
leading to
\begin{align}
X^\pm_{\alpha \beta}&= \frac{1}{2}\left[\left( g_{\rm{FD}}^{\pm +} +  g_{\rm{FD}}^{\pm -}\right)\,\delta_{\alpha\beta} + \frac{(\gamma_5 \slashed{a})_{\alpha\beta}}{\sqrt{-a^2}} \,\left( g_{\rm{FD}}^{\pm +} -  g_{\rm{FD}}^{\pm -}\right) \right]\label{eq:XFDshort},
\end{align}
which represents our novel result.  Note that, by preserving the superscript notation introduced in \EQ{eq:Bolt-short}, the signs in the exponents in \EQ{eq:FD-short} are flipped. This guarantees that the spin-extended Fermi--Dirac distribution reduces to the Boltzmann case in the low-density limit. 

%%%%%%%%%%%%%%%%%%%%%%%%%%%%%
\section{Two-by-two spin density matrix}
\label{sec:2x2}

Contraction of the spinor density matrices $X^\pm$ with the free Dirac spinors ($\urp$ and $\vsp$ with $r,s=1,2$)\footnote{For~more details on our conventions for $\urp$ and $\vsp$, see~\ref{sec:DirBispinors}.}  yields the two-by-two Hermitian spin density matrices 
\begin{align}
    \left[ f^+(x,p) \right]_{rs}  \equiv  \fplusrsxp &=  \frac{1}{2m} \,\ubarrp X^+ \usp, \label{fplusrsxp}  \\
\left[ f^-(x,p) \right]_{rs}  \equiv \fminusrsxp &= - \frac{1}{2m} \, \vbarsp X^- \vrp.\label{fminusrsxp}
\end{align}
Here, the bar denotes the Dirac adjoint. To calculate the right-hand sides of the equations above, we use the normalization conditions $\ubarrp \usp=2m\,\delta_{rs}$ and $\vbarrp \vsp=-2m\,\delta_{rs}$, as well as the identities
\begin{align}
    \ubarrp \gamma_5 \slashed{a}\usp &= 2m \,\bm{a}_* \cdot \sigv_{rs},\\
    \vbarsp \gamma_5 \slashed{a} \vrp &= -2m \,\bm{a}_* \cdot \sigv_{rs},
\end{align}
where $\sigv$ denotes the vector of the three Pauli matrices. The~resulting expressions read
\begin{align}
    f^\pm_{rs}(x,p)&= \frac12 \left[\left( g_{\rm{FD}}^{\pm +} +  g_{\rm{FD}}^{\pm -}\right)\,\delta_{rs} + \frac{\bm{a}_*\cdot\bm{\sigma}_{rs}}{\sqrt{\av_*^2}}\,\left( g_{\rm{FD}}^{\pm +} -  g_{\rm{FD}}^{\pm -}\right)\, \right], \label{eq:frs}
\end{align}
and can be further used to evaluate the mean spin polarization (as defined in~\cite{Florkowski:2017dyn}) 
\begin{align}
    \Pv =  \f{1}{2} \f{ \trt \left[ (f^++f^-) \sigv\right]  }{\trt \left( f^++f^- \right) } = \frac{\bm{a_*}}{2\sqrt{\av_*^2}} \frac{ \displaystyle\sum_{i,j=\pm}j\, g_{\text{FD}}^{ij} }{ \displaystyle \sum_{i,j=\pm}g_{\text{FD}}^{ij}},  \label{eq:meanP}
\end{align}
with the trace taken over the spin indices $r$ and $s$. One can easily verify that the mean spin polarization given \mbox{by \EQ{eq:meanP}} is bounded.

We now turn to the discussion of the direction of the mean spin polarization, using as an example a rigidly rotating fluid in global equilibrium --  also referred to, for brevity, as a vortex in equilibrium. Our results indicate that
\begin{align}
    \bm{P} \, \propto \, -\bm{b}_*,
\end{align}
where~\CITn{Florkowski:2018fap}
\begin{align}
\bv_* = \frac{1}{m} \left(  E_p \, \bv - \pv \times \ev - \f{\pv \cdot \bv}{E_{p} + m} \pv \right). \label{eq:bstar}
\end{align}
In global equilibrium, the spin polarization tensor coincides with the thermal vorticity, \mbox{$\omega_{\mu\nu}=\varpi_{\mu\nu}=-1/2 (\partial_\mu \beta_\nu-\partial_\nu \beta_\mu)$}, and its electric and magnetic components are given by \mbox{$\ev=(0,0,0)$} and \mbox{$\bv = -{\Omega_0}/T_0 \nv_z$}, where ${\Omega_0}$ and $T_0$ are constant parameters, and $\nv_z = (0,0,1)$ is the unit vector that defines the rotation axis. Hence, the polarization vector averaged over particles that form the vortex, $\langle \Pv \rangle$, can be written as
\begin{align}
\langle \Pv \rangle \propto \frac{\Omega_0 \nv_z}{T_0 m} \left(
\langle E_p \rangle -
\left\langle \frac{p_z^2}{E_p+m} \right\rangle \right). \label{eq:Paveraged}
\end{align}
Here, we used cylindrical symmetry to rewrite the second term on the right-hand side. We further notice that \mbox{in \EQn{eq:Paveraged}}, the first term dominates, hence $\langle \Pv \rangle$ points along the $z$-axis, coinciding with the direction of the orbital angular momentum. This is in agreement with previous findings~\CITn{ Becattini:2009wh, Kaparulin:2023fra}.

%%%%%%%%%%%%%%%%%%%%%%%%%%%%%%%%%%%
\section{Macroscopic currents}
\label{sec:currents}

The spinor density matrix \EQn{eq:XFDshort} can be further used to compute the following conserved currents~\cite{DeGroot:1980dk}
\begin{align}
    N^\mu(x)& = \sum_{r=1}^2 \int \dd P  \, p^\mu \left[f^+_{rr}(x,p)-f^-_{rr}(x,p) \right],\label{N^mu_def}\\
    T^{\mu\nu}(x) &= \sum_{r=1}^2 \int \dd P \, p^\mu p^\nu \left[f^+_{rr}(x,p)+f^-_{rr}(x,p) \right],\label{T^munu_def}\\
  \begin{split}  S^{\lambda, \mu\nu}(x) &= \f{1}{2}\sum_{r,s=1}^2 \int \dd P \, p^\lambda  \big[\sigma^{+ \mu\nu}_{sr}(p) f^+_{rs}(x,p) +  \sigma^{- \mu\nu}_{sr} (p) f^-_{rs}(x,p) \big]. \label{S^lmn_def}
  \end{split}
\end{align}
Here, $ N^\mu(x)$ represents the baryon number current, $T^{\mu\nu}(x)$ is the energy-momentum tensor, and  $S^{\lambda, \mu\nu}(x)$ denotes the spin tensor, with \mbox{$\sigma^{+ \mu\nu}_{sr} (p) =1/(2m)\, \ubarsp\, \sigma^{\mu\nu} \urp$,} \mbox{$\sigma^{- \mu\nu}_{sr} (p) =1/(2m)\, \vbarrp\, \sigma^{\mu\nu} \vsp$,} and \mbox{$\sigma^{\mu\nu}=(i/2)[\gamma^\mu,\gamma^\nu]$}.

Equations \EQSMn{N^mu_def}{S^lmn_def} determine the currents in the so-called GLW pseudogauge (the abbreviation refers to the last names of the authors of~\cite{DeGroot:1980dk}). These expressions can be transformed into the canonical and Belinfante--Rosenfeld versions following the procedures outlined in~\cite{Florkowski:2018fap}.

Employing the definitions of the spin density matrices provided in \EQ{eq:frs}, the corresponding expressions for the currents are\footnote{The intermediate steps can be found in~\ref{sec:curr_der}.}
\begin{align}
    N^\mu &= \sum_{i,j=\pm}i\int \dd P\, p^\mu g^{ij}_{\text{FD}},\label{N^mu}\\
    T^{\mu\nu} &=\sum_{i,j=\pm} \int \dd P\, p^\mu  p^\nu g^{ij}_{\text{FD}},\label{T^munu}\\
    S^{\lambda,\mu\nu} &=
    \frac{1}{2m} \sum_{i,j=\pm}j \int \dd P \, p^\lambda  g^{ij}_{\text{FD}} \frac{1}{\sqrt{-a^2}}{\epsilon^{\mu \nu}}_{\alpha \beta}\, a^\alpha p^\beta \label{S^lambdamunu}.
\end{align}
The above-mentioned currents are defined by the integrals, whose convergence requires imposition of a selection criterion, that has been derived and analyzed in details for the spin-extended Boltzmann statistics~\cite{Drogosz:2025ihp,Drogosz:2025iyr}. Since the Fermi–Dirac distribution reduces to the Boltzmann one at high momentum, the same criterion of convergence applies here as well.\par
Based on the studies performed in~\cite{Drogosz:2024gzv} with classical treatment of spin, we introduce a new current
\begin{align}
    \mathcal{N}^\mu =
 - \sum_{i,j=\pm} \int \dd P\, p^\mu \ln(1-g^{ij}_{\text{FD}}) ,
 \label{eq:mathcalN}
\end{align}
which reduces to its Boltzmann counterpart in the low-density regime, $g^{ij}_{\text{FD}}\ll 1$~\cite{Bhadury:2025boe}. Moreover, in traditional hydrodynamics without spin, the \mbox{current~\EQn{eq:mathcalN}} reduces to $\mathcal{N}^\mu=P\beta^\mu$, where $P$ is the local pressure. One can compute the infinitesimal change in~$\mathcal{N}^\mu$ as
\begin{align}
     \dd{\cal N}^\mu &=\frac{\partial{\cal N}^\mu}{\partial \xi}\dd \xi + \frac{\partial {\cal N}^\mu}{\partial \beta_\lambda} \dd \beta_\lambda + \frac{1}{2}\frac{\partial {\cal N}^\mu}{\partial \omega_{\alpha \beta}} \dd \omega_{\alpha \beta}, \label{eq:DcalN}
     \end{align}
with the aid of the relations
     \begin{align}
  \frac{\partial \omega_{\mu \nu}}{\partial \omega_{\alpha \beta}} &= g^\mu_\alpha g^\nu_\beta - g^\nu_\alpha g^\mu_\beta, \qquad     \frac{\partial a^2}{\partial \omega_{\alpha 
    \beta}}=-\frac{1}{m} \epsilon^{\alpha \beta \gamma \delta}a_{\gamma} p_\delta. \label{eq:partials}
\end{align}
The numerical factor of $\,\onehalf$ introduced in \EQ{eq:DcalN} ensures that the spin degrees of freedom are properly counted. This is because, strictly speaking, differentiation should be performed with respect to the independent variables, therefore involving only half of the components of the antisymmetric spin polarization \mbox{tensor~$\omega_{\alpha \beta}$}. 

Using \EQSM{N^mu}{eq:partials}, one can easily verify that the definition given in \EQ{eq:mathcalN} satisfies a generalized Gibbs--Duhem relation
\begin{align}
    \dd{\cal N}^\mu = N^\mu \dd\xi - T^{\mu\lambda} \dd\beta_\lambda + {\frac{1}{2} }S^{\mu, \alpha \beta} \dd\omega_{\alpha\beta}. \label{eq:dNcal}
\end{align}

To supplement our thermodynamic formulation, we define the entropy current for the Fermi--Dirac statistics with spin~\mbox{\cite{Drogosz:2024gzv,Landau:1980mil,doi:10.1142/7396}}
\begin{align}
     S^\mu = -\sum_{i,j=\pm} \int \dd P \dd S\, p^\mu \Big[ g^{ij}_{\text{FD}} \ln g^{ij}_{\text{FD}} + (1-g^{ij}_{\text{FD}})\ln(1-g^{ij}_{\text{FD}})  \Big],
\end{align}
which satisfies a generalized first law of \textit{spin thermodynamics}~\cite{Florkowski:2024bfw},
\begin{align}
    \dd S^\mu = - \xi \dd N^\mu + \beta_\lambda \dd T^{\lambda\mu}- {\frac{1}{2} }\omega_{\alpha\beta} \dd S^{\mu, \alpha \beta}.\label{1stLaw}
\end{align}
Combined with \EQn{eq:dNcal}, Eq.~(\ref{1stLaw}) shows that in local equilibrium, the entropy is conserved as a direct consequence of the conservation of the underlying currents~\cite{Bhadury:2025boe}. We also find
\begin{align}
    S^\mu = - \xi N^\mu + \beta_\lambda  T^{\lambda\mu}- {\frac{1}{2} }\omega_{\alpha\beta}  S^{\mu, \alpha \beta} + {\cal N}^\mu.\label{Smu}
\end{align}

%%%%%%%%%%%%%%%%%%%%%%%%%%%%%%%%%%%%
\section{Generating function for divergence-type treatment}
\label{sec:GenFun}

The divergence-type framework of hydrodynamics~\CITn{LIU1986191, Geroch:1990bw, Peralta-Ramos:2009srp, Peralta-Ramos:2010qdp, Gavassino:2022roi} is a useful tool for analyzing the causality and stability of proposed hydrodynamic theories. As demonstrated in~\cite{Abboud:2025qtg}, perfect spin hydrodynamics can be cast into a divergence type framework, where the primary quantity of interest is the generating function $\chi$, which can be leveraged to obtain all conserved currents. Moreover, it significantly simplifies the analysis of nonlinear causality and equilibrium stability of the theory.

For our case, the generating function takes the form
\begin{align}
 \hspace{-0.23cm}   \chi = \sum_{i,j=\pm}\int \dd P \left[ \frac{1}{2}\left(\ln\frac{1-g^{ij}_{\text{FD}}}{g^{ij}_{\text{FD}}}\right)^2 + \text{Li}_2\left(\frac{g^{ij}_{\text{FD}}-1}{g^{ij}_{\text{FD}}}\right)+\frac{\pi^2}{6}\right],
\end{align}
where $\text{Li}_2(x)$ is the dilogarithm, and the factor $\pi^2/6$ has been added to ensure the convergence of the three-momentum integral. The generating function $\chi$ can then be used to obtain all the relevant currents via the relations
\begin{align}
\begin{split}
    \mathcal{N}^\lambda &=-\frac{\partial\chi}{\partial\beta_\lambda},\qquad \ N^\lambda = -\frac{\partial^2 \chi}{\partial\xi \partial \beta_\lambda}, \qquad
    T^{\lambda\mu} =   \frac{\partial^2 \chi}{\partial \beta_\lambda \partial\beta_\mu}, \qquad S^{\lambda\mu\nu}= -\frac{\partial^2\chi}{\partial\beta_{\lambda}\partial\omega_{\mu\nu}},
    \end{split}
\end{align}
indicating that the theory can indeed be cast into the divergence-type framework. The next natural step is to examine the causality and stability properties of our approach. Positive results for the present framework, as well as for other formulations of perfect spin hydrodynamics (Boltzmann vs. Fermi--Dirac statistics, classical spin treatment vs. spin density description), will be presented in a separate study~\cite{Future}.

%%%%%%%%%%%%%%%%%%%%%%%%%%%%%%%%%%%

\section{Summary and conclusions}
\label{sec:summary}
In this work, a local-equilibrium Wigner function for \mbox{spin-$\onehalf$} particles with dynamical spin degrees of freedom obeying Fermi--Dirac statistics is presented. The proposed function satisfies the proper normalization of the mean spin polarization. Moreover, for a vortex in global equilibrium (where $\omega=\varpi$), the mean spin polarization vector aligns with the direction of the orbital angular momentum. This observation is consistent with known results for rigidly rotating charged fluids in equilibrium. Introducing a new current ($\mathcal{N}^\lambda$), we demonstrated that the conserved currents satisfy generalized thermodynamic relations. The convergence of the integrals defining these currents requires imposition of a selection criterion, as discussed in detail in for the spin-extended Boltzmann statistics~\cite{Drogosz:2025ihp,Drogosz:2025iyr}. Since in the limit of high momentum, the Fermi--Dirac distribution reduces to the Boltzmann case, the same criterion of convergence applies. Finally, we identified a generating function that allows us to cast the underlying theory into a divergence-type framework.

\section*{Acknowledgements}
We thank Wojciech Florkowski for his guidance and clarifying discussions. This work was supported by the National Science Centre (NCN), Poland, Grant No. 2022/47/B/ST2/01372.
\appendix

\appendix
\section{Dirac bispinors}
\label{sec:DirBispinors}
Bispinors $u_r(p)$ and $v_r(p)$ are solutions to the free Dirac equation and are given as 
\begin{equation}
\begin{split}
    u_{r}(p)&=\sqrt{E_{p}+m}\begin{pmatrix}
        \varphi^{(r)}\\[0.5em]
        \frac{\sigv \cdot \pv }{E_{p}+m} \,\, \varphi^{(r)}
    \end{pmatrix}, \qquad
    v_r(p)=\sqrt{E_{p}+m}\begin{pmatrix}
        \frac{\sigv \cdot \pv} {E_{p}+m}\eta^{(r)}\\[0.5em]
        \eta^{(r)}
    \end{pmatrix},
    \end{split}
\end{equation}
where
\beq
    \varphi^{(1)}=\begin{pmatrix}
        1\\
        0
    \end{pmatrix},\qquad \varphi^{(2)}=\begin{pmatrix}
        0\\
        1
    \end{pmatrix},\qquad \eta^{(1)} =\begin{pmatrix}
        0\\
        1
    \end{pmatrix},\qquad \eta^{(2)} =-\begin{pmatrix}
        1\\
        0
    \end{pmatrix}.
\eeq
In the calculations presented below, a bar over the spinor denotes its Dirac adjoint.
\section{Currents from the Wigner function}
\label{sec:curr_der}
\subsection{Baryon number current}
The baryon number current can be computed from the definition in \EQ{N^mu_def} by inserting the proposed spin density matrices \EQSTWOn{fplusrsxp}{fminusrsxp}, yielding
\begin{align}\begin{split}\label{eq:NmuA1} 
N^\mu(x) &=  \int \dd P  \, p^\mu \Bigg[\trt\left(\frac{1}{2m} \,\ubarrp X^+ \usp\right)-\trt \left(- \frac{1}{2m} \, \vbarsp X^- \vrp \right) \Bigg] \\[0.3em]
&= \frac{1}{2m}  \int \dd P  \, p^\mu \left[\trf\left( X^+ (\slashed{p}+m)\right)+\trf \left( X^- (\slashed{p}-m)\right) \right]\\[0.3em]
&=\frac{1}{4m}  \int \dd P  \, p^\mu \left\{\trf\left[\left({g^{++}_{\text{FD}} + g^{+-}_{\text{FD}} } + \frac{\gamma_5\slashed{a}}{\sqrt{-a^2}}\left({g^{++}_{\text{FD}} - g^{+-}_{\text{FD}} }\right) \right) (\slashed{p}+m)\right]\right.\\[0.3em]
&\hspace{0.4cm}\left.+\, \trf \left[\left({g^{-+}_{\text{FD}} + g^{--}_{\text{FD}} } + \frac{\gamma_5\slashed{a}}{\sqrt{-a^2}}\left({g^{-+}_{\text{FD}} - g^{--}_{\text{FD}} }\right) \right) (\slashed{p}-m)\right] \right\}\\[0.3em]
&=\int \dd P p^\mu \left[ {g^{++}_{\text{FD}} + g^{+-}_{\text{FD}} } - \left({g^{-+}_{\text{FD}} + g^{--}_{\text{FD}} }\right) \right]\\[0.3em]
&=\sum_{i,j=\pm}i\int \dd P\, p^\mu g^{ij}_{\text{FD}}.
\end{split}\end{align}
\subsection{Energy momentum tensor}
As was done above, we take \EQ{T^munu_def} for the energy-momentum tensor and rewrite it in terms of traces in spinor indices as
\begin{align}\begin{split}\label{eq:TmunuA2} 
    T^{\mu\nu}(x) &=  \int \dd P  \, p^\mu p^\nu \Bigg[\trt\left(\frac{1}{2m} \,\ubarrp X^+ \usp\right)+\trt \left(- \frac{1}{2m} \, \vbarsp X^- \vrp \right) \Bigg] \\[0.3em]
    &=\frac{1}{2m}  \int \dd P  \, p^\mu p^\nu \Big[\trf\left[X^+ (\slashed{p}+m)\right]-\trf \left[ X^- (\slashed{p}-m)\right] \Big]\\[0.3em]
    &=\frac{1}{4m}  \int \dd P  \, p^\mu p^\nu \left\{\trf\left[ \left({g^{++}_{\text{FD}} + g^{+-}_{\text{FD}} } + \frac{\gamma_5\slashed{a}}{\sqrt{-a^2}}\left({g^{++}_{\text{FD}} - g^{+-}_{\text{FD}} }\right) \right) (\slashed{p}+m)\right]\right.\\[0.3em]
    &\hspace{0.4cm}\left.-\, \trf \left[ \left({g^{-+}_{\text{FD}} + g^{--}_{\text{FD}} } + \frac{\gamma_5\slashed{a}}{\sqrt{-a^2}}\left({g^{-+}_{\text{FD}} - g^{--}_{\text{FD}} }\right) \right) (\slashed{p}-m)\right] \right\}\\[0.3em]
    &=\int \dd P p^\mu p^\nu \left( {g^{++}_{\text{FD}} + g^{+-}_{\text{FD}} }+ {g^{-+}_{\text{FD}} + g^{--}_{\text{FD}} } \right)\\
    &=\sum_{i,j=\pm} \int \dd P\, p^\mu  p^\nu g^{ij}_{\text{FD}}.
\end{split}\end{align}
\subsection{Spin tensor}
The spin tensor in \EQ{S^lmn_def} is obtained from
\begin{align}
    S^{\lambda, \mu\nu}(x) &= \frac{1}{8m^2} \int \dd P \, p^\lambda \Big( \trf \left[\sigma^{\mu\nu} (\slashed{p}+m) X^+ (\slashed{p}+m) \right] -
\trf \left[  \sigma^{\mu\nu} (\slashed{p}-m) X^- (\slashed{p}-m)
  \right]  \Big).
\end{align}
Noting that $[\slashed{p},\gamma_5\slashed{a}]=0$, it follows that $X^\pm$ commutes with $(\slashed{p}+m)$. Hence, we may write
\begin{align}\begin{split}\label{eq:S^lmnA2}
    S^{\lambda, \mu\nu}(x)&=  \frac{1}{4m} \int \dd P \, p^\lambda \left( \trf \left[\sigma^{\mu\nu}  X^+\, (\slashed{p}+m) \right] + \trf \left[  \sigma^{\mu\nu}  X^-\, (\slashed{p}-m)
      \right]  \right) \\[0.3em]
    &=  \frac{1}{8m} \int \dd P \, p^\lambda \left\{ \trf \left[\sigma^{\mu\nu}  \left({g^{++}_{\text{FD}} + g^{+-}_{\text{FD}} } + \frac{\gamma_5\slashed{a}}{\sqrt{-a^2}}\left({g^{++}_{\text{FD}} - g^{+-}_{\text{FD}} }\right) \right)(\slashed{p}+m) \right]\right.\\[0.3em]
    &\hspace{0.4cm}\left. +\, \trf \left[  \sigma^{\mu\nu} \left({g^{-+}_{\text{FD}} + g^{--}_{\text{FD}} } + \frac{\gamma_5\slashed{a}}{\sqrt{-a^2}}\left({g^{-+}_{\text{FD}} - g^{--}_{\text{FD}} }\right) \right)(\slashed{p}-m)
    \right]  \right\} \\[0.3em]
    &=  \frac{1}{8m} \int \dd P \, p^\lambda \left\{ \trf \left[\sigma^{\mu\nu}  \left( \frac{\gamma_5\slashed{a}}{\sqrt{-a^2}}\left({g^{++}_{\text{FD}} - g^{+-}_{\text{FD}} }\right) \right)\,  (\slashed{p}+m) \right]\right.\\[0.3em]
    &\hspace{0.4cm}\left. +\, \trf \left[  \sigma^{\mu\nu}  \left(\frac{\gamma_5\slashed{a}}{\sqrt{-a^2}}\left({g^{-+}_{\text{FD}} - g^{--}_{\text{FD}} }\right) \right)\, (\slashed{p}-m)
    \right]  \right\} \\[0.3em]
    &=\frac{1}{8m}\int \dd P p^\lambda\left( {g^{++}_{\text{FD}} - g^{+-}_{\text{FD}} }+ {g^{-+}_{\text{FD}} - g^{--}_{\text{FD}} } \right)\frac{1}{\sqrt{-a^2}}\trf\left(\sigma^{\mu\nu}\gamma_5\slashed{a}\slashed{p}\right)\\[0.3em]
    &=\frac{1}{2m}\int \dd P p^\lambda\left({g^{++}_{\text{FD}} - g^{+-}_{\text{FD}} }+  {g^{-+}_{\text{FD}} - g^{--}_{\text{FD}} }\right) \frac{1}{\sqrt{-a^2}}{\epsilon^{\mu \nu}}_{\rho \sigma}\, a^\rho p^\sigma\\[0.3em]
    &=\frac{1}{2m} \sum_{i,j=\pm}j \int \dd P p^\lambda  g^{ij}_{\text{FD}} \frac{1}{\sqrt{-a^2}}{\epsilon^{\mu \nu}}_{\rho \sigma}\, a^\rho p^\sigma .
\end{split}\end{align}

\bibliographystyle{elsarticle-num-names} 

\bibliography{ref}



%% For example:

%% \bibitem[Aladro et al.(2015)]{Aladro15} Aladro, R., Martín, S., Riquelme, D., et al. 2015, \aas, 579, A101

%%\end{thebibliography}

\end{document}